\documentclass[prb,aps,slashed,epsfig]{revtex4}
 
\begin{document}

\title{Heavy quarkonia spectra using wave function with gluonic components}

\author{E.A. Bartnik\thanks{correspondence to: bartnik@fuw.edu.pl},~H. Al-Nadary}

\address{Institute of Theoretical Physics, Warsaw University, Ho\.{z}a Street 69, 00-681 Warsaw, Poland}

\begin{abstract}
We calculate the spectra of charmonium and bottomium in an approximation scheme which treats hard
gluons perturbatively while soft gluons are expanded in a set of localized wave functions.
Quark-antiquark and quark-antiquark-gluon sectors are included. Reasonable agreement with 2 parameters
only is found but the spectra are too coulombic. Despite large coupling constant the admixture 
of the quark-antiquark-gluon sector is found to be remarkably small.

\end{abstract}

\pacs{field theory, QCD,quarkonia}

\maketitle

\section{Introduction}

Quantum Chromodynamics (QCD) is a well established theory describing hadron interactions. Due to 
asymptotic freedom its predictions could be verified in high energy collisions. The bound state problem, 
however, remains nearly intractable. The only \emph{ab initio} actually available calculations are obtained 
on the lattice \cite{Bonyakov}. There is, of course, a large body of phenomenological models  \cite{Eduardo} which 
are quite successful in describing the spectra of hadrons. They are invariably based on the assumption that 
a meson is a quark-antiquark state and a baryon a 3 quark state. Here we present an application of a new 
method applicable to the bound state problem in QCD \cite{Bartnik}. It consists in splitting the gluon momenta 
into 2 parts: high momenta, treated perturbatively and low momenta, those gluons are expanded not into plane 
waves but in an orthogonal set of localized functions. As a consequence only a finite number of localized
gluons interact with a heavy quark-antiquark system. High momenta gluons are treated perturbatively within 
one gluon exchange approximation. Here we apply this method to the heavy quarkonia (charmonium and bottomium),
which allows us to use the non relativistic approximation. To find the eigenstates we postulate a variational 
wave function with 2 components: quark-antiquark and quark-antiquark-gluon. After integrating out gluonic 
degrees of freedom we obtain a set of two coupled equations describing quark-antiquark system without and in 
the presence of one gluon. The solution is compared with experimental data. The agreement is not particularly 
good, our effective potentials being too coulombic. However, we use 2 parameters only (quark mass and QCD 
coupling constant). Moreover the inclusion of quark-antiquark-gluon sector improves the agreement.

\section{Zeroth Order Hamiltonian in temporal gauge}

The Lagrange density of QCD is

        \begin{eqnarray}
 {\cal{L}}=-\frac{1}{4}F^{a}_{\mu\nu}F^{\mu\nu}_{a}~+~\overline{\Psi}~(~i D ~-~m~){\Psi}.
      \end{eqnarray}
For the bound state problem in hamiltonian formalism it is convenient to choose a gauge which avoids the 
problem of constraints, the temporal gauge
      \begin{equation}
A_{0}^{a}~=~0.
      \end{equation}
In this gauge we have a 3-vector field ($\vec{A}^a$), its canonically conjugate momenta are simply the 
components of the chromoelectric field 
      \begin{equation}
\dot{A_{i}^{a}}~=~E_{i}^{a},
      \end{equation}
where the Latin index $i$ takes values 1,2,3 and SU(3) gauge index $a$ takes values 1..8. Introducing 
chromomagnetic field
        \begin{eqnarray}
 B^{a}_{i}~=~\varepsilon _{i j k} (\partial_j A_{k}^{a}~+~g f^{a b c} A_{j}^{b} A_{k}^{c})
       \end{eqnarray}
 the gluonic part of the Lagrange density is 
      \begin{equation}
{\cal{L}}~=~\frac{{E^{a}_{i}}~{E^{a}_{i}}}{2} - \frac{{B^{a}_{i}}~{B^{a}_{i}}}{2} 
      \end{equation}
and the corresponding Hamilton density is simply
      \begin{eqnarray}
{\cal{H}}~=~ \frac{{E^{a}_{i}}~{E^{a}_{i}}}{2} + \frac{{B^{a}_{i}}~{B^{a}_{i}}}{2}.
      \end{eqnarray}
Fields $A_{i}^{a}$ and $E_{i}^{a}$ are canonically conjugate and fulfill the commutation relations
        \begin{eqnarray}
[A^{a}_{i}(\vec{x}),E^{b}_{j}(\vec{y})]_{x_{0}=y_{0}}&=&  i\delta_{i j} \delta^{a b}\delta_3(\vec{x}-\vec{y}).   
        \end{eqnarray} 
It should be noted that (6) contains 3 degrees of freedom at each point. The reduction of supernumerary 
degrees of freedom is accomplished by the imposition of Gauss law on the \textbf{state vector}\cite{Susskind}
        \begin{equation}
G^a(x) \mid~ \Psi~ \rangle ~=~ 0.
        \end{equation}
For the quarks we use the non relativistic approximation
       \begin{equation}
 H_{q\bar{q}}~=~\text{Tr} [\frac{1}{2~ m} ~(~p_{1i}~ -~g~ \lambda^{a}_{1} ~A^{a}_{i}~)^2~+~
              \; \frac{1}{2~ m} ~(~p_{2i}~ +~g~\lambda^{a}_{2}~A^{a}_{i}~)^2 ],
        \end{equation}
where $\lambda^a_{1,2}$ are the usual Gell-Mann matrices operating on the color index of the first (second) particle.

\section{High-Low Decomposition}

In order to reduce the problem to a numerically manageable form we have to split our Hamiltonian eqs (6)(9) into
a zeroth order solvable Hamiltonian  and the rest to be treated perturbatively. We split the gluons into 
2 groups:
high momentum (hard) gluons (with ${p}_i > \Lambda$), where $\Lambda$ is the cut off parameter, and low 
momentum (soft) gluons. In high momentum sector we use the conventional approach - the interactions are treated 
as a perturbation. In the lowest order we have free gluons and quarks interacting via \textbf{truncated} coulomb 
potential \cite{Bartnik}. Notice that we have dropped spin-spin and spin-orbit terms.
       \begin{equation}
 V_c~=~\frac{- \alpha_s}{(\vec{p}_1~-~\vec{p}_2)^2},\qquad \mid\vec{p}_1~-~\vec{p}_2\mid> \Lambda ,
        \end{equation}

where $\vec{p}_1$,~$\vec{p}_2$ are the momenta of quark and antiquark respectively and 
$\alpha_s=\frac{4}{3}\alpha$.
For soft gluons we use a different approach. Instead of expanding the field into plane waves, we expand them
into a complete set of \textbf{localized} functions in position space. We have chosen to use 
the Fourier expansion in a cube ~$ -\Lambda> p_i> \Lambda $~in momentum space. In position space 
our wave functions are the product of 3 well known $sinc(x)=sin(x)/x$ functions
       \begin{equation}
e_{\vec{n}}(x,y,z)=(\frac{\Lambda}{\pi})^{\frac{3}{2}}~sinc(n_x \pi -x \Lambda) ~sinc(n_y \pi -y \Lambda)~sinc(n_z \pi -z \Lambda),
        \end{equation}    
where $\vec{n}=(n_x,n_y,n_z), n_x,n_y,n_z$ are integers (positive and negative). 
These functions are distributed on a lattice with lattice spacing $\pi/ \Lambda$. They overlap very little
and, if $\Lambda$ is small enough, the width of the peak will be much larger than the rms radius
of heavy quark-antiquark system. We have found that $\Lambda$= 0.15 - 0.3 GeV was sufficient and the results 
vary very little with $\Lambda$. Thus for soft gluons we have the expansion
       \begin{equation}
A^{a}_{i}(\vec{r})= \sum_{\vec{n}} e_{\vec{n}}(\vec{r}) A^{a}_{i,\vec{n}}
        \end{equation} 
and similarly for the chromoelectric field
              \begin{equation}
E^{a}_{i}(\vec{r})= \sum_{\vec{n}} e_{\vec{n}}(\vec{r}) E^{a}_{i,\vec{n}}
        \end{equation}         
where the $A^{a}_{i}$ and $E^{a}_{i}$ are the operators of localized gluons  and fulfill
        \begin{eqnarray}
  [A^{a}_{i,\vec{n}}~,~E^{b}_{j,\vec{m}}] &=& i\delta_{ij}\delta^{ab} \delta_{3}(\vec{n}-\vec{m}) 
        \end{eqnarray}
Since the wave function of quark-antiquark system "fits" within the wave function at one lattice site
(chosen to have $\vec{n}$~=~(0,0,0) all others gluons will interact very little with quark-antiquark system.
Thus it is reasonable to take into account, in the zeroth approximation, only the gluons with $\vec{n}$~=~(0,0,0).
In fact the quark-antiquark wave function is so narrow that we can approximate~$e_{\vec{n}}(\vec{r})$~by 
its value at the origin. Our zeroth order hamiltonian describes a quark-antiquark pair with a truncated
Coulomb potential interacting with gluons localized at the origin
      \begin{eqnarray}
  H_{0}=   \frac{1}{2~ m} ( \vec{p}_1+g(\frac{\Lambda}{\pi})^{\frac{3}{2}}~\vec{\hat{A}}_1~)^2
   \;+\frac{1}{2~ m} ( \vec{p}_2-g(\frac{\Lambda}{\pi})^{\frac{3}{2}}~\vec{\hat{A}}_2~)^2+{V_c}
   \;+ \frac{{E^a_i}{E^a_i}}{2}~+~\frac{{B^a_i}{B^a_i}}{2}, 
        \end{eqnarray}

where we have used 
\begin{equation}
\lambda^{a}_1 A^{a}_{i}= \vec{\hat{A}_1 }
\end{equation}
and similarly for the antiquark. 
 We have also dropped the index $\vec{n}$~=~(0,0,0).  $B_i^a$ is the chomomagnetic field with derivative 
terms dropped (functions $e_{\vec{n}}(\vec{r})$ are slowly varying within the range of quark-antiquark wave function)
           \begin{eqnarray}
 {B^{a}_{i}}&=& g~\varepsilon _{i j k}~ f^{a b c}~ A_{j}^{b}~ A_{k}^{c}
           \end{eqnarray}
 $f^{a b c}$ are the structure constants of the SU(3) group. We have reduced the problem to a quark-antiquark system 
interacting with $3\times8$=24 gluons localized at the origin. This is numerically tractable by a method 
widely used in atomic and nuclear physics: Ritz variational method. Not all eigenstates of this hamiltonian
are physically acceptable - the wave function must fulfill the Gauss law eq. (8), i.e. the state must
be colorless. After elimination of center of mass motion we get finally
      \begin{eqnarray}
  H_{0}=\frac{1}{2 m}~(~ \vec{p}~+~g~ (\frac{\Lambda}{\pi})^{\frac{3}{2}}~\vec{\hat{A}}_1~)^2 
             \;+\frac{1}{2 m}~(~ \vec{p}~+~g~ (\frac{\Lambda}{\pi})^{\frac{3}{2}}~\vec{\hat{A}}_2~)^2
            \;+{V_c}+ \frac{{E^a_i}{E^a_i}}{2}~+~\frac{{B^a_i}{B^a_i}}{2}~.  
        \end{eqnarray}
where $p$ is the relative momentum of the quark-antiquark.

\section{Variational Ansatz}

The mass of the quark-antiquark system is the energy difference between the energy eigenvalue of Hamiltonian
$ H_{0}$ and the energy of vacuum fluctuations i.e. the lowest energy of gluonic part of  $H_{0}$. We have chosen to 
work in Schr\"odinger  representation with 
        \begin{eqnarray}
 E^{a}_{i}=\frac{1}{i}\partial_o A_{i}^{a}, 
        \end{eqnarray}        
For the vacuum part of the problem we have chosen a variational gaussian Ansatz
        \begin{eqnarray}
{\Psi}_{vac}(\vec{\hat{A}}) ~=~N~ exp \{~-{\beta}_{vac}~\frac{\textrm{Tr}[\vec{\hat{A}}\cdot \vec{\hat{A}}]}{4} \},
        \end{eqnarray}
$\beta$ being the variational parameter. We obtain
        \begin{eqnarray}
{\beta}_{vac}&=& \frac{4~ 2^{\frac{1}{3}} \alpha^{\frac{1}{3}} \Lambda}{(3\pi)^\frac{2}{3}}=0.168~\textrm{GeV},\\      
E_{vac} &=&\frac{12~{6^{\frac{1}{3}}}~{\alpha^{\frac{1}{3}}}~\Lambda }{\pi^{\frac{2}{3}}}=1.511~\textrm{GeV}.
        \end{eqnarray}

For the quark-antiquark system we postulate a 2 component wave function
       \begin{eqnarray}
{\Psi}(~\vec{p}~,\vec{\hat{A}}~)~=~{\delta}_{ij}~~{f}_1(\vec{p})~~exp \{~-{\beta}~~\frac{\textrm{Tr}[\vec{\hat{A}}\cdot \vec{\hat{A}}]}{4} \}
     \;+{f_2}(~\vec{p}~)~\frac{\vec{p}~\vec{\hat{A}}}{p}~ 
     \;~exp \{~-{\beta}~~\frac{\textrm{Tr}[\vec{\hat{A}}\cdot \vec{\hat{A}}]}{4} \}
       \end{eqnarray}
The first part describes a quark antiquark system with no gluons (gaussian i.e. lowest energy state 
of the oscillator) while the second part represents a quark-antiquark-gluon system (the
gluonic part is proportional to the first excited state of an oscillator). Both parts are colorless.
The variation is to be done
with respect to $\beta$, ${f}_1(\vec{p})$ and ${f}_2(\vec{p})$ with the condition that the norm of the 
wave function ${\Psi}(~\vec{p}~,\vec{\hat{A}}~)$ be one. After solving the variational problem for 
${f}_1(\vec{p})$ and ${f}_2(\vec{p})$ and integrating out gluonic degrees of freedom we have a system 
of two coupled equations
         \begin{eqnarray}
           \left( \begin{array}{cc}
     	            {h}_{11}&{h}_{12}\\
     	            {h}_{21}&{h}_{22}
          \end{array} \right)
      \left( \begin{array}{cc}
	            {f}_1(\vec{p})\\
	            {f}_2(\vec{p})
          \end{array} \right)~=~\mathcal{E}
          \left( \begin{array}{cc}
	  	            {f}_1(\vec{p})\\
	  	            {f}_2(\vec{p})
          \end{array} \right), 
        \end{eqnarray}  
where

        \begin{eqnarray}
 {h}_{11} &=& \frac{p^2}{2~ \mu}~+~
            \;\widetilde{V_c}(\vec{r})~+~
            \;\frac{4~ g^2 ~\Lambda^3}{\pi^3 ~\beta~ \mu}~+~6~\beta~+~\frac{32~ g^2 ~\Lambda^3}{3~ \pi^3~ \beta^2}\\       
 {h}_{22} &=& \frac{p^2}{2~ \mu}~+~
            \;\widetilde{V_c}(\vec{r})~+~
            \;\frac{31~ g^2 ~\Lambda^3}{6~\pi^3 ~\beta~ \mu}~+~
            \;\frac{13~\beta}{2}~+~\frac{112~ g^2 ~\Lambda^3}{9~ \pi^3~ \beta^2}\\
 {h}_{12} &=& \big(\frac{8~g~\Lambda^{\frac{3}{2}}}{3~\beta~\mu~\pi^{\frac{3}{2}}}\big)~{p}\\
 {h}_{21} &=& \big(\frac{g~\Lambda^{\frac{3}{2}}}{\mu~\pi^{\frac{3}{2}}}\big)~{p}.
        \end{eqnarray}.
        
Notice that diagonal terms differ by a constant only and that the coupling terms are proportional to $p$.
This set of equations was solved numerically by expanding ${f}_1(\vec{p})$ and ${f}_2(\vec{p})$ in the
basis of Laguerre polynomials. We found that a basis of 16 polynomials was sufficient.

\section{Results and discussion}

We have calculated spectra of charmonium and bottomium. In both cases we used 
       \begin{equation}
\Lambda~=~0.15~\textrm{GeV}
        \end{equation}    
In our \emph{ab initio} calculation we have 2 free parameters only: the mass of the quark and the coupling
constant. For charmonium we found
       \begin{eqnarray}
       {m}_{c}&=&1.8604~\textrm{GeV},\\
       {{\alpha}_s}_c&=&0.973,
       \end{eqnarray}
by fitting the 2 lowest L=0 states. Since we have no spin-spin interaction in our non relativistic approximation
${}^{1}\textrm{S}_{0}$ and ${}^{3}\textrm{S}_{1}$ are degenerate, we used weighted  the average \cite{Lichtenberg}. 
All the experimental data are from ref \cite{Data}.
\\
The results are summarized
in Table 1 and Table 2 and shown in Fig. 1 and Fig. 2. We show the results for a calculation with quark-antiquark
sector only (${\mathcal{E}}_{q\bar{q}}$), with quark-antiquark-gluon only (${\mathcal{E}}_{q\bar{q}g}$) and full coupled 
channels (${\mathcal{E}}_{full}$). We consider the agreement reasonable considering that only 2 parameters were used. 
In our approximation the system is too coulombic but the
inclusion of quark-antiquark-gluon sector improves the agreement with experimental data. To our surprise the admixture
of quark-antiquark-gluon sector is very small; for the ground state we have 
       \begin{eqnarray}
       {p}_{q\bar{q}g}&=&0.0017
       \end{eqnarray}
 
 despite using a large value for ${{\alpha}_s}_c \approx 1$. This could be an explanation why 
 phenomenological models with constituent quarks only are so successful  \cite{Eduardo}.
 For bottomium the results are similar. We find
        \begin{eqnarray}
        {m}_{b}&=&5.1896~\textrm{GeV},\\
        {{\alpha}_{s}}_{b}&=&0.66114.
        \end{eqnarray}

  Notice that the coupling constant is 30\% lower. Again the spectrum is too close to a purely coulombic one
  but the inclusion of quark-antiquark-gluon sector improves the agreement (Table 3-4 and Fig 3-4). The admixture
 of quark-antiquark-gluon sector is very small
          \begin{eqnarray}
          {p}_{q\bar{q}g}&=&0.0004.
           \end{eqnarray}

 The main feature of this calculation is the surpisingly small admixture of quark-antiquark-gluon sector. However
 the long range part of the effective potential is clearly wrong. The obvious improvement would be to take into account 
 space variation of gluon wave function and the inclusion of gluons located on neighbour sites on the lattice. While 
 the computation of spectra would be vastly more complicated it is certainly feasible. In conclusion it is seen 
 that it is possible to perform bound state calculations within QCD and improve them step by step.

\newpage
\begin{center}
\begin{tabular}{|r||c|c|l|}\hline
 Exp.~& ~${\mathcal{E}}_{q\bar{q}}$~&~${\mathcal{E}}_{q\bar{q}g}$~&~${\mathcal{E}}_{full}$~ \\ \hline
\hline
3.068~&~3.0955~&~3.8565~&~3.068~ \\ \hline
3.663~&~3.6785~&~3.9469~&~3.663~ \\ \hline
4.040~&~3.7687~&~3.9659~&~3.761~ \\ \hline
4.415~&~3.7977~&~4.0106~&~3.785~ \\ \hline
\end{tabular}
\qquad \qquad\qquad \qquad 
\begin{tabular}{|r||c|c|l|}\hline
 Exp.~& ~${\mathcal{E}}_{q\bar{q}}$~&~${\mathcal{E}}_{q\bar{q}g}$~&~${\mathcal{E}}_{full}$~ \\ \hline
\hline
3.494~&~3.6797~&~3.9513~&~3.656~ \\ \hline
     ~&~3.770 ~&~3.9615~&~3.762~ \\ \hline
     ~&~3.7891~&~3.9925~&~3.779~ \\ \hline
     ~&~3.8338~&~4.0437~&~3.812~ \\ \hline
\end{tabular}
\end{center}
\begin{center}
\mbox{Table 1~:~ $S$ levels in $c\bar{c}$~(in GeV)} \qquad \qquad \qquad  \mbox{Table 2~:~ $P$~levels in $c\bar{c}$~(in GeV)}
\end{center}


\begin{center}
\begin{tabular}{|r||c|c|l|}\hline
 Exp.~& ~${\mathcal{E}}_{q\bar{q}}$~&~${\mathcal{E}}_{q\bar{q}g}$~&~${\mathcal{E}}_{full}$~ \\ \hline
\hline
9.460 ~&~9.469 ~&~10.381~&~9.460~ \\ \hline
10.023~&~10.224~&~10.517~&~10.218~ \\ \hline  
10.355~&~10.361~&~10.57 ~&~10.355~ \\ \hline
10.580~&~10.427~&~10.648~&~10.385~ \\ \hline
10.865~&~10.555~&~10.805~&~10.421~ \\ \hline 
11.019~&~10.863~&~11.141~&~10.519~ \\ \hline
\end{tabular}
\qquad \qquad\qquad \qquad 
\begin{tabular}{|r||c|c|l|}\hline
 Exp.~& ~${\mathcal{E}}_{q\bar{q}}$~&~${\mathcal{E}}_{q\bar{q}g}$~&~${\mathcal{E}}_{full}$~ \\ \hline
\hline
9.888 ~&~10.224~&~10.518~&~10.215~ \\ \hline
10.253~&~10.361~&~10.564~&~10.356~ \\ \hline 
      ~&~10.414~&~10.613~&~10.408~ \\ \hline
      ~&~10.492~&~10.703~&~10.479~ \\ \hline 
      ~&~10.649~&~10.868~&~10.523~ \\ \hline
\end{tabular}
\end{center}
\begin{center}
\mbox{Table 3~: $S$ levels in $b\bar{b}$~(in GeV)}\qquad \qquad\qquad \qquad \mbox{Table 4~: $P$ levels in $b\bar{b}$~(in GeV)}
\end{center}


{}

\end{document}